\documentstyle[12pt]{article}
\newcommand{\be}{\begin{equation}}
\newcommand{\ee}{\end{equation}}

\newcommand{\ensec}[1]{\section{#1} \setcounter{equation}{0}}

\newcommand{\gS}{\Sigma}
\newcommand{\jj}[1]{\mbox{\boldmath $#1$}}

\title{\sc BLACK HOLE ENTROPY AND PHYSICS AT PLANCKIAN SCALES}

\vspace{.5cm}
\author{{\sc valeri frolov} \vspace{.3cm} \\ {\small {\em CIAR
Cosmology
Program;  Theoretical Physics Institute,}} \\  {\small {\em
University of
Alberta, Edmonton, Canada T6G 2J1}}\thanks{E-mail address:
frolov@phys.ualberta.ca}}
\date{}
\begin{document}
\maketitle

\vspace{.3cm}

\ensec{Black-Hole Entropy}
According to the  thermodynamical analogy in  black hole physics,
the
entropy of a black hole  in the Einstein theory of gravity is
 \begin{equation}\label{0}
\jj{S}^{BH} =\jj{A}_H
/(4l_{\mbox{\scriptsize{P}}}^2),
\end{equation}
where   $\jj{A}_H$   is   the   area   of   a   black   hole
surface  and
$l_{\,\mbox{\scriptsize{P}}}=
(\hbar G/c^3)^{1/2}$   is   the   Planck length
\cite{Beke:72,Beke:73}.
In  black hole physics the Bekenstein-Hawking entropy $\jj{S}^{BH}$
plays
essentially the same role as in the usual thermodynamics.  In
particular it
allows  to estimate what part of the internal energy of a black hole
can be
transformed into work. Four laws of black
hole physics that form the basis in the thermodynamical analogy  were
formulated in \cite{BaCaHa:73}.  According to this analogy the
entropy $\jj{S}$
is defined by the response of the free energy $\jj{F}$ of the system
containing
a black hole to the change of  its temperature:
\begin{equation}\label{1}
d\jj{F} = -\jj{S} dT .
\end{equation}
The generalized second law \cite{Beke:72,Beke:73,Beke:74} (see also
\cite{ThPrMa:86,NoFr:89,Wald:92,FrPa:93} and references therein)
implies that
when a black hole is a part of the thermodynamical system the total
entropy
(i.e. the sum of the entropy of a black hole and the entropy of the
surrounding
matter) does not decrease.

The Euclidean approach \cite{HaHa:76,GiHa:76, Hawk:79} provides a
natural way
to derive black hole thermodynamical properties. Doing a Wick's
rotation
$t\rightarrow -i\tau$ in the Schwarzschild metric one gets  the
metric with the
Euclidean signature. The corresponding manifold with the Euclidean
metric is
regular if and only if the imagine time $\tau$ is periodic and the
period is
$2\pi/\kappa$, where $\kappa$ is the surface gravity of the black
hole. The
corresponding regular Euclidean metric is known as the
Gibbons-Hawking
instanton. The period $2\pi/\kappa$ of the imagine time $\tau$ is
naturally
identified with the inverse temperature of the black hole, while the
Euclidean
action, calculated for the Gibbons-Hawking instanton, is directly
related with
its free energy.  York and collaborators \cite{York:86, BrBrWhYo:90}
made an
important observation that the formal derivation of the black hole
entropy
using the Euclidean approach requires  certain modifications  for its
consistency. Namely,  to get a well defined canonical ensemble, one
needs to
consider a black hole in a box of finite size and fix the
corresponding
thermodynamical quantities (temperature) at the boundary. Box filled
with
thermal radiation and a black hole in it in thermal equilibrium with
radiation
is thermodynamically stable if the radius of a spherical box $r$ is
less than
$3M$.  If one changes the temperature of the box, the mass of a black
hole
inside the box also changes, as  required by the conditions of the
thermal
equilibrium. As the result of this process the free energy $\jj{F}$
of the
system is changed.  One can use the relation (\ref{1}) and single out
the
contribution to the entropy due to the black hole. In the classical
approximation the so defined thermodynamical entropy  coincides with
the
Bekenstein-Hawking entropy $\jj{S}^{BH}$.

The simple relation between the thermodynamical entropy of a black
hole and its
surface area is characteristic for Einstein theory. The definition of
the black
hole entropy can be generalized to non-Einstein versions of
gravitational
theory, provided they allow existence of black holes. Wald
\cite{Wald:93}
showed that in the general case the  entropy of a black hole is
defined by
N\"{o}ther charge related with the Killing vector. (For application
and
developing this idea, see \cite{JaMy:93,  IyWa:94, JaKaMy:94}.)
Recently the
interest to the problem of black hole entropy was increased by
observation
that the entropy of extremal black hole might vanish
\cite{GiKa:95,HaHoRo:95}.

The success of the thermodynamical analogy in black hole physics
allows one to
hope that this analogy may be  even deeper and it is possible to
develop
statistical-mechanical foundation of black hole thermodynamics. It is
worthwhile to remind that the
thermodynamical and statistical-me\-cha\-ni\-cal definitions of the
entropy are
 logically different. {\it Thermodynamical entropy} $\jj{S}^{TD}$ is
defined by
the response of the free energy $\jj{F}$ of a system to the change of
its
temperature:
\begin{equation}\label{1a}
d\jj{F} = -\jj{S}^{TD} dT .
\end{equation}
(This definition applied to a black hole determines its
Bekenstein-Hawking
entropy.)

{\it Statistical-mechanical entropy} $\jj{S}^{SM}$ is defined as
\begin{equation}\label{2}
\jj{S}^{SM}=-\mbox{Tr}(\hat{\jj{\rho}} \ln \hat{\jj{\rho}} ) ,
\end{equation}
where $\hat{\jj{\rho}}$ is the density matrix describing the internal
state of
the system under consideration.  It is also possible to introduce the
{\it
informational entropy} $\jj{S}^I$ by counting different possibilities
to
prepare a system in a  final state with given macroscopical
parameters from
different initial states
\begin{equation}\label{3}
\jj{S}^I =-\sum_{n}{\jj{p}_n \ln\jj{p}_n },
\end{equation}
with $\jj{p}_n$ being the probabilities of different initial states.
In standard case all three definitions give the same answer.

Is the analogy between black holes thermodynamics and the 'standard'
thermodynamics complete? Are there  internal degrees of freedom of a
black hole
 responsible for its entropy?  Is it possible to apply the
statistical-mechanical and informational definitions of the entropy
to black
holes and how are they related with the Bekenstein-Hawking entropy?
These are
the questions that are to be answered.

Historically  first attempts of the statistical-me\-cha\-nical
foundation of
the entropy of a black hole were connected with the informational
approach
\cite{Beke:73,ZuTh:85}. According to this approach the black hole
entropy is
interpreted as "the logarithm of the number of quantum mechanically
distinct
ways that the hole could have been made"\cite{ThPrMa:86, ZuTh:85}.
The so
defined informational entropy of a black hole is simply related to
the amount
of information lost by stretching the horizon, and as was shown by
Thorne and
Zurek it is equal to the Bekenstein-Hawking entropy \cite{ThPrMa:86,
ZuTh:85}.
Quite interesting results relating informational and
statistical-mechanical
entropies can be obtained in a special model proposed by Bekenstein
and
Mukhanov \cite{Mukh:86, BeMu:95}.  According to this model a black
hole is
identified with a system having discrete internal states, so that an
absorption
or emission of a particle by a black hole is accompanied by the
transition from
one state to another. In such  model the mass of a black hole is
quantized\footnote{Another approach to the quantization of the mass
of a black
hole can be found in \cite{Bere:90}.}. Unfortunately the physical
origin of
internal degrees of freedom of a black hole in the model and their
discreteness
 is not derived but postulated.

The  dynamical origin of the entropy of a black hole and the relation
between
the statistical-mecha\-ni\-cal and Bekenstein-Hawking entropy have
remained
unclear. In the present talk I describe some recent results obtained
in this
direction.

\ensec{Dynamical Degrees of Freedom of a Black Hole}

The problem of the dynamical origin of the black hole entropy was
intensively
discussed recently.  The proposed basic idea   is to relate the
dynamical
degrees of freedom of a black hole with its quantum excitations. This
idea has
different realizations\footnote{
For recent review of the problem of the dynamical origin of the
entropy of a
black hole, see  also \cite{Beke:94}.
}.

In the framework of the {\it membrane paradigm} the dynamical degrees
of
freedom are identified with  different possible states of  thermal
atmosphere
of a black hole, while the entropy of a black hole is identified with
the
amount of information about the state of  thermal atmosphere, which
is lost by
stretching the horizon\cite{ThPrMa:86, ZuTh:85}.

In his {\it brick wall model}  't Hooft \cite{Hoof:85} proposed to
consider a
mirror-like boundary, located outside a black hole at the close
distance to its
horizon.  He assumed that outside the boundary there exist thermal
radiation
with temperature equal to the Hawking temperature. He has shown that
the
entropy of such thermal radiation is of the same order of magnitude
as the
Bekenstein-Hawking entropy, provided the distance of the mirror-like
boundary
from the horizon is chosen to be of order of Planck's length.

Bombelli {\it et al} \cite{BoKoLeSo:86} attracted attention to the
fact that
even in a flat spacetime in a Minkowski vacuum state one can obtain a
non-vanishing entropy if one restricts his observations to the
spatially
bounded part of the space. The corresponding {\it entanglement
entropy} arises
due to the presence of correlations of those modes of  zero-point
fluctuations
which are propagating in the vicinity of the boundary of chosen
spatial region.
It was proposed to relate black hole entropy with the entanglement
entropy
related with the presence of the horizon \cite{BoKoLeSo:86, Sred:93}

We (with Igor Novikov) arrived to the similar idea independently by
analyzing
the gedanken experiment proposed in our earlier paper
\cite{FrNo:93a}. Namely
we assumed that there exist a traversable wormhole, and its mouths
are freely
falling into a black hole. If one of the mouths crosses the
gravitational
radius earlier than the other, then rays passing through the first
mouth can
escape from the region lying inside the gravitational radius. Such
rays would
go through the wormhole  and enter the outside region though the
second mouth.
As the result  during the period of time when the first of the mouth
is inside
and the other mouth is outside the gravitational radius the surface
area of the
 horizon  decreases.  If we assume that the black hole entropy is
related with
the surface area of a black hole, then the only possibility to escape
contradictions with the second law is to assume that during such
process the
decrease of entropy is related with the possibility to get access to
some new
information concerning black hole internal states. At first sight it
looks like
a puzzle. We know that (at least in the classical General Relativity)
a black
hole  at late time is completely specified by finite number of
parameters. For
a non-rotating uncharged black hole one need to know only one
parameter (mass
$M$) to describe all its properties. It is true not only for the
exterior where
this property is the consequence of no-hair theorems, but also for
the black
hole interior  \cite{DoNo:78}. The reason why it is impossible for an
isolated
black hole at late time to have non-trivial classical states in its
interior in
the vicinity of the horizon is basically the same as for the exterior
regions.
These states might be excited only if a collapsing body emits the
pulse of
fields or particles immediately after it crosses the event horizon.
Due to
red-shift effect the energy of the emitted pulse must be
exponentially large in
order to reach the late time region with any reasonable energy. After
short
time (say $100 M$) after the formation of a black hole it is
virtually
impossible because it requires the energy of emission much greater
than the
black hole mass $M$.

In quantum physics the situation is completely different. The above
mentioned
puzzle can be solved if we remember that any quantum field has
zero-point
fluctuations.  To analyze  states of a  quantum field it is
convenient  to use
its decomposition into modes. Besides the positive-frequency modes
which have
positive energy, there exist also positive-frequency modes with
negative
energy. In a non-rotating uncharged black holes such modes can
propagate only
inside the horizon where the Killing vector  used to define the
energy is
spacelike. It is possible to show that at late time for any regular
initial
state of the field these states are thermally excited and the
corresponding
temperature coincides with the black hole temperature $T_H=(8\pi
M)^{-1}$.
(For more details  see, e.g.\cite{FrNo:93b}.) In principle by using a
traversable wormhole described above one can get information
concerning these
internal modes propagating near the horizon and change their states.

In the framework of this {\it dynamical-black-hole-interior} model we
proposed
to identify the dynamical degrees of freedom of a black hole with the
internal
modes of all physical fields.  The set of the fields must include the
gravitational one. It can be shown that for any chosen field  the
number of the
modes infinitely grows as one considers the regions located closer
and closer
to the horizon. For this reason   the contribution of a field to the
statistical-mechanical  entropy of a black hole calculated by
counting the
internal modes of a black hole is formally divergent. In order to
make it
finite one might restrict himself by considering only those modes
which are
located at the proper distance from the horizon greater than some
chosen value
$l$. For this choice of the cut-off the contribution of a field to
the
statistical-mechanical entropy of a black hole is
\be \label{2.1}
\jj{S}^{SM}=\alpha{\jj{A}\over l^2},
\ee
where $\jj{A}$ is the surface area of a black hole, and a
dimensionless
parameter $\alpha$ depends on the type of the field.

\ensec{No-Boundary Wave Function of a Black Hole}
The calculation of the black hole entropy in the
dynamical-black-hole-interior
 model is made by counting the number of thermally excited internal
modes
existing at given moment of time (more accurately on the chosen
spacelike
surface, crossing the horizon). This calculations can be simplified
by using
the following trick. Consider an eternal version of a black hole,
i.e. an
eternal black hole with the same mass $M$ as the original black hole
formed as
the result of collapse. At late time the geometry of both holes are
identical.
One can trace back in time all the perturbations, propagating at late
time in
the geometry of the eternal version of a black hole . As the result
one can
relate   perturbations at late time in a spacetime of a real black
hole with
initial data on the Einstein-Rosen bridge (spatial slice of $t=$const
) of the
eternal black hole geometry (see Fig.~1).
\begin{figure}
\label{f1}
\let\picnaturalsize=N
\def\picsize{7cm}
\def\picfilename{f1.eps}
\ifx\nopictures Y\else{\ifx\epsfloaded Y\else\input epsf \fi
\let\epsfloaded=Y
\centerline{\ifx\picnaturalsize N\epsfxsize \picsize\fi
\epsfbox{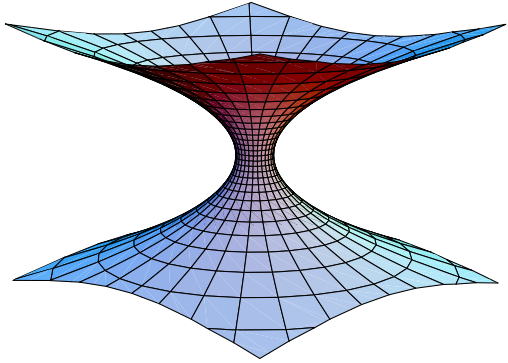}}}\fi
\vspace{-2.5cm}
\caption[f1]{Embedding diagram for a two-dimensional
$\theta=\mbox{const}$
section of the Einstein-Rosen bridge.}
\end{figure}

We denote the 3-surface of the Einstein-Rosen bridge by $\jj{\gS}$.
This
surface has the topology $S^2\times R^1$. The 2-surface $\jj{S}$ of
the horizon
$r=2M$ splits it into two isometric parts: 'external' $\jj{\gS}_{+}$
and
'internal' $\jj{\gS}_{-}$. In a spacetime of the eternal black hole
the Killing
vector $\xi$ which is used to define energy is future directed on
$\jj{\gS}_+$
and past directed on $\jj{\gS}_-$. For this reason initial date
having a
support located on $\jj{\gS}_+$ correspond to the field
configurations having
positive energy, while the energy of the field configurations with
the initial
data on $\jj{\gS}_-$ possess negative energy. The former describes
external
degrees of freedom of a black hole, while the latter describes the
internal
ones. The set of fields representing the degrees of freedom of a
black hole
contains the gravitational perturbations. For given initial values of
fields
and gravitational perturbations on $\jj{\gS}$ the gravitational
constraint
equations determine the deformation of the 3-geometry of the
Einstein-Rosen
bridge (see Fig.~2). We shall use the notion 'deformation' in order
to describe
 not only deformed geometry of the Einstein-Rosen bridge, but also
the physical
fields on it.  By using this terminology we can say that  the states
of a black
hole at late time are uniquely related with deformations of the
Einstein-Rosen
bridge.
\begin{figure}
\label{f2}
\let\picnaturalsize=N
\def\picsize{7cm}
\def\picfilename{f2.eps}
\ifx\nopictures Y\else{\ifx\epsfloaded Y\else\input epsf \fi
\let\epsfloaded=Y
\centerline{\ifx\picnaturalsize N\epsfxsize \picsize\fi
\epsfbox{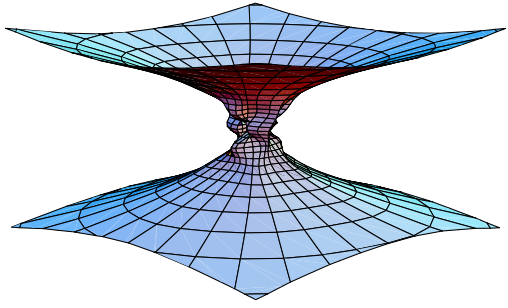}}}\fi
\vspace{-2.5cm}
\caption[f2]{Deformation of the Einstein-Rosen bridge.}
\end{figure}

It was proposed in \cite{BaFrZe:94} to introduce a wave function of a
black
hole as the functional over the space of deformations of the
Einstein-Rosen
bridge. In this approach the wave function of a black hole depends on
data
located on both parts of the Einstein-Rosen bridge: an external
$\mbox{\boldmath $\Sigma$}_{+}$ (external degrees of freedom) and  an
inner
$\mbox{\boldmath $\Sigma$}_{-}$  one (internal degrees of freedom).

Certainly there exist infinite number of different wavefunctions of a
black
hole.  Our aim is to get a useful tool for the description of the
canonical
ensemble of black holes inside the cavity restricted by a spherical
boundary of
the radius $r_B$ and with fixed inverse temperature $\beta$ on it.
For this
reason only very special wavefunctions  will be important for us.
Here we present a modified version of the no-boundary approach of
Ref.\cite{BaFrZe:94} which is analogous to the 'no-boundary ansatz'
in quantum
cosmology \cite{HaHa:83}. This ansatz singles out a set of
no-boundary
wavefunctions   which  is convenient for our purpose.

Instead of the complete Einstein-Rosen bridge we  consider its part
$\jj{\gS}^{'}$ lying  between two spherical 2-dimensional boundaries
$\jj{S}_{\pm}$ located from both sides of $\jj{S}$ at the radius
$r=r_B$.
$\jj{\gS}^{'}$ has the topology $S^2\times I$, where $I$ is the unit
interval
$[0,1]$. We denote by $\mbox{\boldmath $M$}_{\beta}$ a Euclidean
manifold with
a boundary $\partial\mbox{\boldmath $M$}_{\beta}$, which consists of
two parts:
$\jj{\gS}^{'}$ and another 3-surface $\jj{\gS}^B$ with the same
topology
$S^2\times I$, which intersects $\jj{\gS}^{'}$ at $\jj{S}_+$ and
$\jj{S}_-$,
and which represents the Euclidean evolution of the external boundary
$B$.

We define the no-boundary wavefunction depending on one parameter
$\beta$ by
the following path integral
	\begin{eqnarray}
	\mbox{\boldmath $\Psi$}_{\beta}(^3\!g(\mbox{\boldmath $x$}),
	\varphi(\mbox{\boldmath $x$}))=
	\int {\cal D} \,{}^4\!g \ {\cal D}\mbox{\boldmath $\phi$}\
	{\rm e}^{\!\!\phantom{0}^
	{\textstyle -\mbox{\boldmath $I$}
	[\,{}^4\!g ,\mbox{\boldmath $\phi$}\,]}} .
\label{3.1}
	\end{eqnarray}
Here $\mbox{\boldmath $I$}[\,{}^4\!g
,\mbox{\boldmath $\phi$}\,]$ is the Euclidean gravitational action.
The
integral is taken over   Euclidean 4-geometries and matter-field
configurations on a  spacetime $\mbox{\boldmath $M$}_{\beta}$ with a
boundary
$\partial\mbox{\boldmath $M$}_{\beta}\equiv \jj{\gS}^{'}\cup
\jj{\gS}^{B}$. The
integration variables are subject to the
conditions
	\[(^3\!g(\mbox{\boldmath $x$}),\,
	\varphi(\mbox{\boldmath $x$})),\;\;
	\mbox{\boldmath $x$}
	\in\partial\mbox{\boldmath $M$}_{\beta},\]
-- the collection of 3-geometry and boundary
matter fields  on $\partial\mbox{\boldmath $M$}_{\beta}$, which are
just the
argument
of the wavefunction (\ref{3.1}).

We assume that the 3-metric on the boundary is of the form
\be
ds^2_{\gS^B}=d\tau^2+r_B^2d\omega^2 +\ldots,
\hspace{1cm}\tau\in(-\beta/4,\beta/4), \label{n1}
\ee
\be
ds^2_{\gS^{'}}=(1-r_+/r)^{-1}dr^2 +r^2d\omega^2 +\ldots,
\hspace{1cm}r\in
[r_+,r_B), \label{n2}
\ee
where $r_+\equiv 2M$, $d\omega^2$ is the line element on a unit
sphere, and
dots indicate omitted terms describing perturbations of the metric.

If $(g_0,\phi_0)$ is a point of the extremum of the action $\jj{I}$,
then we
can write
\be
g=g_0+\tilde{g},\hspace{1cm}\phi=\phi_0+\tilde{\phi},\label{n3}
\ee
\be \label{n4}
\jj{I}[g,\phi]=\jj{I}_0[g_0,\phi_0]+\jj{I}_2[\tilde{g},\tilde{\phi}]
+\ldots .
\ee
In accordance with this decomposition the no-boundary wavefunction
(\ref{3.1})
in the quasiclassical approximation reads
\be \label{n5}
\mbox{\boldmath $\Psi$}_{\beta}(^3\!g(\mbox{\boldmath
$x$}),\varphi(\mbox{\boldmath $x$}))=\mbox{\boldmath
$\Psi$}_{\beta}^0(^3\!g_0(\mbox{\boldmath
$x$}),\varphi_0(\mbox{\boldmath
$x$}))\ \  \mbox{\boldmath
$\Psi$}^1_{\beta}(^3\!\tilde{g}(\mbox{\boldmath
$x$}),\tilde{\varphi}(\mbox{\boldmath $x$})) ,
\ee
where
\be
\mbox{\boldmath $\Psi$}_{\beta}^0(^3\!g_0(\mbox{\boldmath
$x$}),\varphi_0(\mbox{\boldmath $x$}))=\mbox{e}^{\textstyle
-\jj{I}{_0}[g_0,\phi_0]} ,
\ee
is a classical (tree-level) contribution, and
\be \label{n6}
\mbox{\boldmath $\Psi$}^1_{\beta}(^3\!\tilde{g}(\mbox{\boldmath
$x$}),\tilde{\varphi}(\mbox{\boldmath $x$}))=
	\int {\cal D} \,{}^4\!\tilde{g} \ {\cal
D}\tilde{\mbox{\boldmath ${\phi}$}}\
	{\rm e}^{\!\!\phantom{0}^
	{\textstyle  -\mbox{\boldmath $I$}_{2}
	[\,{}^4\!\tilde{g} ,\tilde{\mbox{\boldmath ${\phi}$}}\,]}} .
\ee
is a one-loop part.

We consider a theory for which $\phi_0=0$, so that $g_0$ is a
solution of the
vacuum Einstein equations. The corresponding Euclidean solution is a
part
$\jj{M}_{\beta}$ of the Gibbons-Hawking instanton (see Fig.~3), i.e.
the
Euclidean Schwarzschild solution
\be
ds^2=Fd\tau^2+F^{-1}dr^2+d\omega^2, \hspace{1cm}F=1-r_+/r ,
\ee
with $\tau\in (-{1\over 4}\beta_{\infty},{1\over 4}\beta_{\infty})$,
where
$\beta_{\infty}=(F(r_B))^{-1/2}\beta$ is the inverse temperature at
infinity.
(For a special choice of $\beta_{\infty}=8\pi M$ this part is a half
of the
instanton.)
\begin{figure}
\label{f3}
\let\picnaturalsize=N
\def\picsize{5cm}
\def\picfilename{f3.eps}
\ifx\nopictures Y\else{\ifx\epsfloaded Y\else\input epsf \fi
\let\epsfloaded=Y
\centerline{\ifx\picnaturalsize N\epsfxsize \picsize\fi
\epsfbox{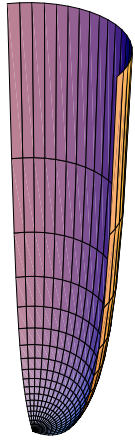}}}\fi
\caption[f3]{Part $\jj{M}_{\beta}$  of the Gibbons-Hawking instanton,
which
gives the main contribution into the no-boundary wavefunction
$\jj{\Psi}_{\beta}$ in the quasiclassical approximation.}
\end{figure}

To calculate the Euclidean Einstein-Hilbert action
\be\label{10}
\jj{I}_0[g_0]=-{1\over{16\pi}}\int_{\jj{M}_{\beta}}R\sqrt{g}
d^4x-{1\over{8\pi}}
\int_{\partial\jj{M}_{\beta}}K\sqrt{h}d^3x -{1\over{8\pi}}
\int_{\partial\jj{M}_{\beta}}K_0\sqrt{h}d^3x
\ee
for $\jj{M}_{\beta}$ we note that for the vacuum solution $R=0$, so
that only
the surface terms of the action contribute. The calculation of the
trace of the
extrinsic curvature $K$ for $\jj{\gS}^B$ is straightforward and gives
\be
K={1\over 2}{r_+\over r^2_B}{1\over \sqrt{F(r_B)}}+{2\sqrt{F(r_B)}
\over r_B} \
{}.
\ee
By using this expression we get
\be
{1\over{8\pi}}\int_{\jj{\gS}^{B}}K\sqrt{h}d^3x={1\over 2}\beta r_B
\sqrt{F(r_B)}+{\beta r_+\over 8\sqrt{F(r_B)}}\ .
\ee
The extrinsic curvature vanishes identically everywhere on the
boundary
$\jj{\gS}^{'}$ except the point $r=r_+$,  where it has $\delta$-type
singularity. The corresponding contribution is
\be
{1\over{8\pi}}\int_{\jj{\gS}^{B}}K\sqrt{h}d^3x={\pi\over 2}r^2_+
\left(
1-{\beta \over {4r_+\sqrt{F(r_B)}}}\right) .
\ee
There is a well known ambiguity in the term
${1\over{8\pi}}\int_{\partial\jj{M}_{\beta}}K_0\sqrt{h}d^3x$ which is
to be
subtracted and which depends on the choice of a reference space. Here
we fix
this ambiguity simply by subtracting from
${1\over{8\pi}}\int_{\partial\jj{M}_{\beta}}K\sqrt{h}d^3x$ its value
for $M=0$.
Finally we have
\be
\jj{I}_0[g_0]={1\over 2}\beta r_B \left( 1-\sqrt{F(r_B)}\right)
-{\pi\over
2}r^2_+ \ .
\ee

Now we calculate the one-loop contribution to the no-boundary
wavefunction of a
black hole. First of all we note that each of the fields (including
the
gravitational perturbations) give independent contribution to
$\jj{I}_2$. It
means that $\jj{\Psi}^1_{\beta}$ is a product of wavefunctions
$\jj{\Psi}^1_{\beta}[\varphi]$ depending on only one particular type
of field
$\varphi$. (We remind that $\phi_0=0$ and hence the value $\varphi$
on the
boundary coincides with its perturbation $\tilde{\phi}$.) The
Gaussian integral
(\ref{n6}) in the definition of $\jj{\Psi}^1_{\beta}[\varphi]$ can be
easily
calculated. Really, let us denote by $\phi(\varphi)$ a solution of
field
equations for the action $\jj{I}_2[\phi]$ obeying the boundary
conditions
$\phi\left|_{\partial \jj{M}_{\beta}}\right.=\varphi$. Then
\be
\jj{\Psi}^1_{\beta}(\varphi)=C\mbox{e}^{\textstyle
-\jj{I}_2[\varphi]}.
\ee
We assume that $\varphi$ on $\jj{\gS}^B$ does not depend on $\tau$.
In this
case a solution $\phi(\varphi)$ obeying boundary conditions
	\begin{eqnarray}
	\phi\,(x)\,\Big|_{\textstyle\,
	\mbox{\boldmath $\Sigma$}_{\pm}}
	\equiv\phi\,(\pm \beta_{\infty}/4,{\mbox{\boldmath $x$}})
	=\varphi_{\pm}({\mbox{\boldmath $x$}}),
\label{n4.3}
	\end{eqnarray}
can be written as a decomposition
	\begin{eqnarray}
	\phi (\tau,{\mbox{\boldmath
        $x$}})=\sum_{\lambda}\Big\{\,\varphi_{\lambda,+}
	u_{\lambda,-}(\tau,{\mbox{\boldmath $x$}})+
	\varphi_{\lambda,-}
	u_{\lambda,+}(\tau,{\mbox{\boldmath $x$}})\,\Big\}
\label{n4.4}
	\end{eqnarray}
in the basis functions of the field equation
	\begin{eqnarray}
	&&u_{\lambda,\pm}(\tau,{\mbox{\boldmath $x$}})=
	\frac{{\rm sinh}\,(\beta_{\infty}/4\mp\tau)\,\omega}
	{{\rm
sinh}\,(\beta_{\infty}/2)}\,R_{\lambda}({\mbox{\boldmath $x$}}),
           \label{n4.5}
	\end{eqnarray}
where  $R_{\lambda}({\mbox{\boldmath
$x$}})$ is a complete  set of spatial harmonics on $\jj{M}_{\beta}$
with a
chosen boundary conditions on $\jj{\gS}^B$.
The coefficients $\varphi_{\lambda,\pm}$ in
(\ref{n4.4}) are just the decomposition coefficients of the fields
(\ref{n4.3})
in the basis of spatial harmonics
	\begin{eqnarray} \label{n4.10}
	\varphi_{\pm}({\mbox{\boldmath $x$}})=
	\sum_{\lambda} \varphi_{\lambda,\pm}
	R_{\lambda}({\mbox{\boldmath $x$}}).
	\end{eqnarray}

Substituting (\ref{n4.4}) into $\jj{I}_2[\varphi]$, integrating by
parts with
respect to
the
Euclidean time and
spatial coordinates and taking into account the equations of motion,
one
finds that the Euclidean action reduces to the following quadratic
form in
$\varphi_{\lambda,\pm}$:
	\begin{eqnarray}
	\mbox{\boldmath $I$}_2[\varphi_{+},\varphi_{-}]=
	\sum_{\lambda}\left
	\{\,\frac{\omega_{\lambda}\,{\rm cosh}
	(\beta_{\infty}\,\omega_{\lambda}/2)}
	{2\ {\rm sinh}(\beta_{\infty}\,\omega_{\lambda}/2)}
	\,(\varphi_{\lambda,+}^2+\varphi_{\lambda,-}^2)
	-\frac{\omega_{\lambda}}{{\rm sinh}(\beta_{\infty}\,
	\omega_{\lambda}/2)}
	\,\varphi_{\lambda,+}\varphi_{\lambda,-}\!\right\}.
\label{n4.11}
	\end{eqnarray}
This action is a sum of Euclidean actions for quantum oscillators of
frequency
$\omega_{\lambda}$ for the interval $\beta_{\infty}$ of the Euclidean
time with
the initial value of its amplitude $\varphi_-$ and the final value
$\varphi_+$.

To summarize we obtain the following expression for the no-boundary
wavefunction  of a black hole in the semiclassical approximation
\begin{eqnarray}
	\jj{\Psi}_{\beta}[M, \varphi_{+},\varphi_{-}]=
	N\,{\rm e}^{\!\!\phantom{0}^{
	{\textstyle {  -1/2\beta r_B [1-(1-2M/r_B)^{1/2}]+2\pi M^2\!
-
\sum\mbox{\boldmath $I$}_{2}[\varphi_{+},\varphi_{-}]}}}},
     \label{3.14}
	\end{eqnarray}
Here $N$ is a normalization constant. The symbol of summation in the
exponent
indicates that the additional a summation over all physical fields
must be
done. The square of this wavefunction gives the probability to find a
given
configuration in the state determined by the parameter $\beta$. For
large $M$
($M\gg m_{\mbox{P}}$) this probability is a sharp peak with width
$\approx
m_{\mbox{P}}$ located near the value of $M=M_{\beta}\equiv
\beta_{\infty}/8\pi$. For $\beta_{\infty}=8\pi M$ and $r_B\rightarrow
\infty$
this wavefunction coincides with a no-boundary wavefunction obtained
in
\cite{BaFrZe:94}.

For fixed $M$ the density matrix for internal variables $\varphi_-$
of a black
hole is defined as
\be
\hat{\jj{\rho}}_{\beta}[\varphi_-,\varphi_-^{'}]=\int D\varphi_{+}
\jj{\Psi}_{\beta}[\varphi_+,\varphi_-]
\jj{\Psi}_{\beta}[\varphi_+,\varphi_-^{'}] ,
\ee
and it is of the form
\begin{eqnarray}
	\hat{\jj{\rho}}[\varphi_-,\varphi_-^{'}]=P^{'}\,{\rm
e}^{\textstyle{-\tilde{\jj{I}}_{2}[\,\varphi_-,\varphi_-^{'}\,]}} ,
     \label{3.14c}
	\end{eqnarray}
where $\tilde{\mbox{\boldmath ${I}$}}_2$ is given by the expression
(\ref{n4.11}) with $\beta$ changed by $2\beta$. It is easy to show
that
\be\label{3.a8}
\hat{\jj{\rho}}[\varphi_-,\varphi_-^{'}]=P^{''}\langle \varphi_-|{\rm
e}^{\textstyle{-\beta \hat{\mbox{\boldmath
${H}$}}}}|\varphi_-^{'}\rangle ,
\ee
where $P^{''}$ is a normalization constant and

$\hat{\mbox{\boldmath ${H}$}}$
is the Hamiltonian of  free fields $\varphi$ propagating on the
Schwarzschild
background. The statistical-mechanical  entropy $\jj{S}^{SM}$ of a
black hole
obtained by using this density matrix coincides with the expression
(\ref{2.1}).

The statistical-mechanical  entropy $\jj{S}^{SM}$  in this as well as
other
'dynamical' approaches possesses the following main properties: (1)
$\jj{S}^{SM}\sim \jj{A}$, where $\jj{A}$ is the surface area of a
black hole;
(2) $\jj{S}^{SM}$ is divergent and requires regularization:
$\jj{S}^{SM}\sim
\jj{A}/l^2$, where $l$ is the cut-off parameter; (3) $\jj{S}^{SM}$
depends on
the number of fields, which exist in nature; (4)~$\jj{S}^{SM}\sim
\jj{S}^{BH}$
for $l\sim l_{\mbox{P}}$ .

The following two problems are of  importance: (1) What is the
relation between
the statistical-mechanical  entropy $\jj{S}^{SM}$ introduced by
counting the
internal degrees of freedom of a black hole and its thermodynamical
entropy
$\jj{S}^{TD}$? In particular how to explain the universality of the
Bekenstein-Hawking entropy $\jj{S}^{BH}$, while $\jj{S}^{SM}$ is not
universal
and depends on the number of fields? (2) The formal expression for
the
statistical-mechanical entropy $\jj{S}^{SM}$ contains the Planck
scale cut-off.
Does it mean that by studying the thermodynamical properties of black
holes we
can obtain certain conclusions concerning physics at Planckian
scales?

In what follows we shall try to clarify these questions.

\ensec{Renormalized Effective Action and Free Energy}
The complete information concerning the canonical ensemble of black
holes with
a given inverse temperature $\beta$ at the boundary is contained in
the
partition function $\jj{Z}(\beta)$ given by the Euclidean path
integral
\cite{Hawk:79}
\begin{equation}\label{2.2}
\jj{Z}(\beta)=\int D[\jj{\phi}]\exp (-\jj{I} [\jj{\phi}]) .
\end{equation}
Here the integration is taken over all fields including the
gravitational one
that  are real on the Euclidean section and are periodic in the
imaginary time
coordinate $\tau$ with period $\beta$. The quantity $\jj{\phi}$ is
understood
as the collective variable describing  the fields. In particular it
contains
the gravitational field. Here $D[\phi]$ is the measure of the space
of fields
$\jj{\phi}$ and $\jj{I}_E$ is the Euclidean action of the field
configuration.
The action $\jj{I}_E$ includes the Euclidean Einstein action. The
state of the
system is determined by the choice of the boundary conditions on the
metrics
that one integrates over. For the canonical ensemble for the
gravitational
fields inside a spherical box of radius $r_B$ at temperature $T$ one
must
integrate over all the metrics inside $r_B$ which are periodically
identified
in the imaginary time direction with period $\beta=T^{-1}$. Such a
partition
function must describe in particular the canonical ensemble of black
holes.
The partition function $\jj{Z}$ is related with the effective action
$\jj{\Gamma}=-\ln \jj{Z}$ and with the free energy
$\jj{F}=\beta^{-1}\jj{\Gamma}=-\beta^{-1}\ln \jj{Z}$.

By using the stationary-phase approximation one gets
\begin{equation}\label{2.3}
\beta \jj{F}\equiv - \ln \jj{Z} = \jj{I}[\jj{\phi}_0] - \ln \jj{Z}_1
+\ldots  .
\end{equation}
Here $\jj{\phi}_0$ is the (generally speaking, complex)  solution of
classical
field equations for action $\jj{I}[\phi]$ obeying the required
periodicity and
boundary conditions. Besides the tree-level contribution
$\jj{I}[\jj{\phi}_0]$,
the expression (\ref{2.3}) includes also one-loop corrections  $\ln
\jj{Z}_1$,
connected with the contributions of the fields perturbations on the
background
$\jj{\phi}_0$, as well as higher order terms in loops expansion,
denoted by
$(\ldots )$. The  one-loop contribution of a field $\jj{\phi}$ can be
written
as follows    $\ln \jj{Z}_1 =-{1 \over 2}\mbox{Tr} \ln (-\jj{D})$,
where
$\jj{D}$ is the  field operator  for the field $\jj{\phi}$ inside
the  box
$r_B$.  The one-loop contribution contains divergences and required
the
renormalization. In order to be able to absorb these divergences in
the
renormalization of the coefficients of the initial classical action
we chose
the latter in the form
\be \label{b}
\jj{I}_{cl}=\int d^4x \sqrt{g}\jj{L} ,
\ee
\be
\jj{L}=\left[ -\frac{\Lambda_B}{8\pi G_B} -\frac{R}{16\pi G_B}
+c^1_B R^2 +c^2_B R_{\mu\nu}^2 +c_B^3
R_{\alpha\beta\mu\nu}^2 \right] .
\ee

By using heat-kernel representation for $\ln \jj{Z}_1$ one can write
\be \label{a.1}
-{1\over 2}\ln \mbox{det} (-\jj{D})={1\over 2}
\int_{\delta^2}^{\infty}{ds\over s}\mbox{Tr} K(s) ,
\ee
where $K(s)$ is the heat-kernel of the operator $D$ which has the
following
Schwinger-DeWitt expansion
\be
K(s)=e^{-s\jj{D}}={1\over 16\pi^2 s^2}\sum a_n s^n , s\rightarrow 0 .
\ee
For the particular case of a scalar massless field
\be
a_0 =1, \hspace{1cm}a_1=(1/6 -\xi)R,
\ee
\be
a_2 ={1\over 180}R_{\alpha\beta\mu\nu}^2 -
{1\over 180}R_{\mu\nu}^2 -{1\over
6}({1\over 5}-\xi)\Box R +{1\over 2}({1\over 6}-\xi )^2 R^2
\ee
By substituting this expansion into (\ref{a.1}) one can conclude that
the
one-loop contribution $\jj{\Gamma}_1$ to the effective action can be
written in
the form
\be
\jj{\Gamma}_1=\jj{\Gamma}_1^{div}+\jj{\Gamma}_1^{fin} .
\ee
where
\be
\jj{\Gamma}_1^{div}= -{1\over 32\pi^2}\int d^4x \sqrt{g}
\left[ {a_0\over
2\delta^4}+{a_1\over \delta^2}-2a_2\ln (\delta)\right] .
\ee
The divergent part of the one-loop effective action has the same
structure as
the initial classical action (\ref{b}) and hence one can write
\be \label{3.a}
\jj{\Gamma}= \jj{\Gamma}_{cl}^{ren}+ \jj{\Gamma}_1^{ren},
\ee
\be
\jj{\Gamma}_1^{ren}=\jj{\Gamma}_1-\jj{\Gamma}_1^{div}
=\jj{\Gamma}_1^{fin} .
\ee
Here $\jj{\Gamma}_{cl}^{ren}$ is identical to the initial classical
action with
the only change that all the bare coefficients $\Lambda_B$, $G_B$,
and $c^i_B$
are substituted by their renormalized versions $\Lambda_{ren}$,
$G_{ren}$, and
$c^i_{ren}$
\be
{\Lambda_{ren}\over G_{ren}}={\Lambda_{B}\over G_{B}}+
{1\over 8 \pi \delta^4} ,
\ee
\be
{1\over G_{ren}}={1\over G_{B}}+{1\over 2\pi \delta^2}
({1\over 6}-\xi) ,
\ee
\be
c^i_{ren}=c^i_B+\alpha^i\ln \delta .
\ee
We shall refer to (\ref{3.a}) as to the loop expansion of the
renormalized
effective action. After multiplying the the renormalized effective
action by
$\beta^{-1}$ we get the expansion for the renormalized free energy.

The effective action $\jj{\Gamma}$ contains the complete information
about the
system under consideration. In particular the variation of
$\jj{\Gamma}$ with
respect to the metric provides one with the equations for the quantum
average
metric $\bar{g}=\langle g\rangle$:
\be\label{3.a2}
{\delta\jj{\Gamma}\over{\delta{\bar g}}}=0 .
\ee
One usually assumes that quantum corrections are small and solves
this equation
perturbatively:
\be
\bar g=g_{cl}+\delta g ,
\ee
where $g_{cl}$ is a solution of the classical equations. At this
point we need
to make an important remark. In principle, there exist two
possibilities:
either to begin with the solution of the classical equations for the
action
(\ref{b}), or its renormalized version $\jj{\Gamma}^{ren}_{cl}$,
which is
written in terms of the renormalized constants. One usually assumes
that the
renormalized values of  $\Lambda_{ren}$ and $c^i_{ren}$ vanish
$\Lambda_{ren}=c^i_{ren}=0$. It means that in general case their
initial values
were not vanishing unless one is dealing with some special type of
theory (e.g.
assuming supersimmetry). In other words the global properties of the
solutions
for $\jj{I}_{cl}$ and $\jj{\Gamma}^{ren}_{cl}$ are generally
different. So  to
provide the condition that $\delta g$ is small,  one is to begin with
the
metric $g_{cl}$ that is an extremum of $\jj{\Gamma}^{ren}_{cl}$
\be\label{3.a4}
{\delta\jj{\Gamma}^{ren}_{cl}\over{\delta{g}_{cl}}}=0 .
\ee
We assume that  the renormalization of the coupling constants in the
classical
action is made from the very beginning and we shall assume that the
'classical'
field $g_{cl}$ is a solution of the equation  (\ref{3.a4}). In our
case
$g_{cl}$ is the Euclidean black hole metric, while the metric
$\bar{g}$
describes the Gibbons-Hawking instanton deformed due to the presence
of quantum
corrections to the metric. The quantity $\jj{\Gamma}[\bar{g}]$ being
expressed
as the function of boundary conditions ($\beta$ and $r_B$) specifies
the
thermodynamical properties of a black hole.

\ensec{Thermodynamical Entropy}
For the above described canonical ensemble of gravitational fields
the leading
tree-level  contribution to  the renormalized effective action is
given by the
Euclidean gravitational action for the Euclidean black hole solution
(the
Gibbons-Hawking instanton).
\begin{figure}
\label{f4}
\let\picnaturalsize=N
\def\picsize{5cm}
\def\picfilename{f4.eps}
\ifx\nopictures Y\else{\ifx\epsfloaded Y\else\input epsf \fi
\let\epsfloaded=Y
\centerline{\ifx\picnaturalsize N\epsfxsize \picsize\fi
\epsfbox{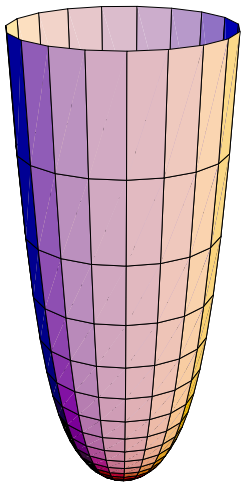}}}\fi
\caption[f4]{Embedding diagram for a two-dimensional $(\tau ,r)$
section of the
Gibbons-Hawking instanton. Regularity condition at the Euclidean
horizon
$r=r_+$ requires $\beta=\beta_H\equiv 8\pi G_{ren}M$. }
\end{figure}

Under the assumption that $\Lambda_{ren}=0$ and $c^i_{ren}=0$ the
tree-level
contribution  to the renormalized  free energy of the black hole  is
\cite{York:86,BrBrWhYo:90}
\begin{equation}\label{2.4}
\jj{F}_0^{ren}\equiv \beta^{-1} \jj{\Gamma}^{ren}_{cl}[g_{cl}]= r_B
\left(
1-\sqrt{1-r_+ /r_B}\right) -\pi
r_+^2\beta^{-1} .
\end{equation}
Here  $r_+ =2G_{ren}M$ is the gravitational radius of a black hole of
mass $M$,
which for a given temperature $\beta^{-1}$ at the boundary $r_B$ is
defined by
the relation
$\beta=4\pi r_+ (1-r_+ /r_B )^{1/2}$. According to  definition the
thermodynamical entropy of a black hole $\jj{S}^{TD}_0$ is determined
by the
response of the free energy of a system including a black hole to the
change of
the temperature. One can easily verify that
\begin{equation}\label{2.5}
\jj{S}^{TD}_0=-{d{\jj{F}_0^{ren}} \over dT}\equiv \beta^2
{d{\jj{F}_0^{ren}}
\over d\beta}=\frac{\jj{A}_H}{4l_{\mbox{\scriptsize{P}}}^2},
\end{equation}
and hence it coincides with the Bekenstein-Hawking expression
$\jj{S}^{BH}$.
(It is assumed that $r_+$ in $\jj{F}_0^{ren}$ is expressed in terms
of $\beta$
and $r_B$ before differentiation with respect to $\beta$.) One-loop
contribution in Eq.(\ref{2.3}) describes quantum correction to the
entropy of a
black hole as well as the entropy of thermal radiation in its
exterior. The
latter evidently depends on the radius $r_B$ of the boundary.
Since the Euclidean black hole background is regular the
corresponding
contribution $\jj{F}^{ren}_1$  is finite.  For this reason the
quantum
corrections to the Bekenstein-Hawking entropy $4\pi M^2$ are also
finite. They
are small unless the mass of a black hole $M$ is comparable with the
Planckian
mass \cite{York:94,Whit:90,Whit:95}. Due to the presence of the
conformal
anomalies one might expect that the leading one-loop corrections to
$S^{TD}$
are of the order $\ln M$ (see, e.g. \cite{Furs:95}).

\ensec{Statistical-Mechanical Entropy}
The derivation of the thermodynamical entropy of a black hole
requires the {\it
on-shell} calculations. It means that one uses only a  regular
Euclidean metric
that is  solution of the field equations (\ref{3.a2}). The discussion
of the
relation of the thermodynamical and statistical-mechanical entropy of
a black
hole requires {\it off-shell} calculations. The reason for this is
quite simple
and can be explained, for example, by using the approach based on the
no-boundary wavefunction of BFZ \cite{BaFrZe:94}. The matrix elements
in the
$|\varphi_-\rangle$ basis of the operator $\hat{\rho}\ln \hat{\rho}$
which
enters the definition of the statistical-mechanical entropy (\ref{2})
can be
obtained by partially differentiating (\ref{3.a8}) with respect to
$\beta$. On
the other hand the Hamiltonian $\hat{\jj{H}}$ depends on the
black-hole
geometry, and hence on the mass $M$ of a black hole. That is why to
obtain the
expression for the statistical-mechanical entropy one needs to be
able to use
$\beta$ and $M$ as independent parameters.
\begin{figure}
\label{f5}
\let\picnaturalsize=N
\def\picsize{5.0cm}
\def\picfilename{f5.eps}
\ifx\nopictures Y\else{\ifx\epsfloaded Y\else\input epsf \fi
\let\epsfloaded=Y
\centerline{\ifx\picnaturalsize N\epsfxsize \picsize\fi
\epsfbox{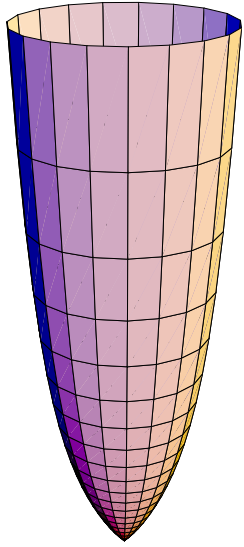}}}\fi
\caption[f5]{Embedding diagram for a two-dimensional $(\tau ,r)$
section of the
Euclidean black hole with cone-like singularity  ($\beta\ne
\beta_H\equiv 8\pi
G_{ren}M$).}
\end{figure}

Strictly speaking for $\beta\ne \beta_H\equiv 8\pi M$ there are no
regular
Euclidean solutions with the Euclidean black-hole topology $R^2\times
S^2$.
Such solutions can be obtained only if one exclude a horizon sphere
$S^2$. One
can also consider the spacetime with included horizon, provided the
curvature
has there $\delta$-like behavior corresponding to the cone-like
singularity of
the metric. For $\beta=\beta_H$ the singularity dissapears. In order
to be able
to discuss the statistical-mechanical entropy and its relation to the
thermodynamical entropy one must generalize the calculation of the
one-loop
contribution to the renormalized free energy to the case of spaces
with
cone-like singularity. The new feature which arises is that the
corresponding
renormalized one-loop corrections might contain new type of
divergence, which
is directly connected with the presence of cone singularity. In order
to make
the answer finite one might introduce spatial cut-off in the volume
integrals
near the cone-singularity. It is convenient to restrict the
integration by
some proper distance $l$ from singularity. This cut-off was present
in the 't
Hooft's 'brick wall model' \cite{Hoof:85}. In the model with
'dynamical
black-hole-interior' the presence of such a cut-off was connected
with the
quantum fluctuations of the horizon \cite{FrNo:93b}. Similar cut-off
naturally
arises in the string theory \cite{VeSa:88}.
The renormalized  one-loop contribution $F^{ren}_1$ to the free
energy is of
the form
\be\label{4.a1}
F^{ren}_1=F^{ren}_1[\beta, \beta_H,\varepsilon] ,
\ee
where $\varepsilon=(l/2G_{ren}M)^2$ is the dimensionless cut-off
parameter. For
 $\varepsilon\rightarrow0$ and $\beta\ne\beta_H$ the one-loop free
energy
$F^{ren}_1$ is divergent $F^{ren}_1\sim \varepsilon^{-1}f(\beta,
\beta_H)$. For
$\beta=\beta_H$ the divergence dissappears, so that
$F^{ren}_1[\beta_H,
\beta_H,0]$ is finite. This quantity is directly related with quantum
(one-loop) corrections to the thermodynamical entropy of a black hole
\be\label{4.a2}
\jj{S}^{TD}_1=\beta^2_H  {d\over d\beta_H}\jj{F}^{ren}_1[\beta_H,
\beta_H,0]\equiv \left[\beta^2 {\partial \jj{F}^{ren}_1\over \partial
\beta}+\beta^2_H{\partial \jj{F}^{ren}_1\over \partial
\beta_H}\right]
_{\beta=\beta_H} .
\ee

The expression (\ref{4.a1}) allows one to get the
statistical-mechanical
entropy $\jj{S}^{SM}$.
Dowker and Kennedy \cite{DoKe:78} and  Allen \cite{Alle:86} made an
important
observation that
\be
\jj{F}^{ren}_1=\jj{F}^{ren}_{vac} +\jj{F}^{ren}_{therm},
\hspace{1cm} {\partial
\jj{F}^{ren}_{vac}\over \partial \beta}=0 ,
\ee
\be
\jj{F}^{ren}_{therm}=-\beta^{-1}\ln \mbox {Tr}\left[ \mbox{e}^{-\beta
\hat{\jj{H}}}\right] =\ln [\sum \exp (-\beta E_n )] ,
\ee
where $E_n$ is the energy (eigenvalue of the Hamiltonian
$\hat{\jj{H}}$ of the
field
$\varphi$).
By using the expansion in eigenfunctions one can obtain
\be\label{5.a1}
\jj{F}^{ren}_{therm}=\sum_{\lambda}{f(\beta\omega_{\lambda}})
=\int{d\omega
N(\omega|\beta_H,\varepsilon)}f(\beta\omega) .
\ee
$f(\beta\omega)=\beta^{-1}\ln [1-\exp (-\beta\omega )]$ is free
energy of an
oscillator of
frequency $\omega$ at inverse temperature $\beta$, and $N(\omega
|\beta_H)$ is the density of number of states at the given energy
$\omega$ in a
spacetime of a
black hole of mass $M$. This density of number of states diverges. In
order to
make it finite we introduced the  cut-off $\varepsilon$. We include
$\varepsilon$ as the argument of $N$ in order to remind about this.
The
expression (\ref{5.a1}) is usually a starting point for 'brick wall'
model.

The statistical-mechanical entropy $\jj{S}^{SM}$ is
\be\label{4.a3}
\jj{S}^{SM}=\left[ {\partial \jj{F}^{ren}_1\over \partial
\beta}\right]
_{\beta_H}
=\left[ {\partial \jj{F}^{ren}_{therm}\over \partial \beta}\right]
_{\beta_H}
\ee
\be
=\int{d\omega
N(\omega|\beta_H,\varepsilon)}s(\beta\omega)
\ee
Here $s (\beta\omega ) = {\beta\omega } / ( e^{\beta\omega }-1) -\ln
(1-
e^{-\beta\omega })$ is the entropy of a quantum oscillator of
frequency
$\omega$ with inverse temperature $\beta$.  $\jj{S}^{SM}$ is
divergent in the
limit $\varepsilon\rightarrow 0$. The divergence  is directly related
with the
divergency of the density of number of states located in the narrow
region in
the vicinity of the horizon.

By comparing the expressions (\ref{4.a2}) and (\ref{4.a3}) we can
conclude that
$\jj{S}^{TD}$ and $\jj{S}^{SM}$ differs from one another. It happens
for the
following two reasons: (1) Vacuum polarization ($\jj{F}^{ren}_{vac}$)
depends
on $M$ and hence  on $\beta_H$; (2)  $d/d\beta$ does not commute with
Tr-operation. In general case one gets
\be\label{4.a4}
\jj{S}^{TD}_1= \jj{S}^{SM}+\Delta \jj{S}.
\ee
In the limit $\varepsilon\rightarrow 0$  $\Delta \jj{S}\ne 0$ is also
divergent, but $\jj{S}^{TD}_1$ remains finite and (for $M\gg
m_{\mbox{P}}$)
small.
The relation (\ref{4.a4})  provides explanation of the entropy
renormalization
procedure by Thorne and Zurek \cite{ZuTh:85}.

Fursaev and Solodukhin \cite{FuSo:94, So:95, FuSo:95} recently
proposed another
approach for off-shell calculations of thermodynamical
characteristics of a
black hole. Namely, instead of cutting-off the vicinity of a cone
singularity,
they proposed to calculate the effective action directly on a
spacetime with
cone-like singularity. In order to make this mathematically well
defined one
might at first  consider a manifold with the topology $R^2\times S^2$
which is
smooth at the fixed-point sphere (horizon) and differs from the cone
metric
only in the very narrow region of size $l$ near the horizon. The
effective
action must be considered as the function of $l$ and the parameter
$l$ must be
finally put to zero. The one-loop correction to the free energy
$\jj{F}^{cone}_1$ can be divergent in this limit, but it is possible
to show
that the divergence is proportional to $(\beta-\beta_H)^2$. That is
why the
corresponding contribution  $\jj{S}^{cone}_1=\left[ \beta^2 \partial
\jj{F}^{cone}_1/ \partial \beta\right]_{\beta=\beta_H}$ to the
entropy is
finite. This off-shell approach  gives the same expression for the
thermodynamical entropy and might be considered as  useful tools for
such
calculations. The relation of this off-shell entropy
$\jj{S}^{cone}_1$ to the
statistical-mechanical entropy $\jj{S}^{SM}$ is not  clear. Among
other
approaches to  the calculation of the entropy we mention also the
approach
based on the Pauli-Villars regularization \cite{Myer:95}.

\ensec{Black Hole Thermodynamics and Physics at Plan\-ckian Scales}
The expression for the statistical-mechanical entropy (2.1) requires
cut-off.
The value $l$ of the cut-off parameter  is of order of the Planckian
length..
Does it mean that  thermodynamical characteristics of a black hole
for their
understanding require the knowledge of physics at Planckian scales?

When we are discussing black hole solutions and their properties we
use
gravitational equations. The coupling  constants in these equations
are assumed
to coincide with 'observable' values. Due to the existence of
ultraviolet
divergencies the observable coupling constants differ from their
initial bare
values. Any procedure which gives sense to this renormalization
procedure
finally must deal with the problems of physics at Planckian scale.
But in this
sense black hole physics does not differ  from the usual Newtonian
theory.
Calculations in the Newtonian theory also use the renormalized
('observable')
gravitational constant, and hence in order to derive the same results
in the
framework of quantum gravity beginning from some initial background
theory one
must pass through all the complications connected with the
renormalization
procedure and redefining the coupling constants. One can do it from
very
beginning or develop more complicated scheme and made all the
renormalizations
only at the end. The same is true also for black holes. Besides this
in the
case of black hole there are  situations when quantum gravity becomes
really
important.   It is  well known that the final stage of a black hole
evaporation
as well as the structure near singularity inside a black hole for
their
consideration require quantum gravity. Quantum gravity might be also
useful for
study of small quantum corrections to black hole characteristics. But
these
corrections can essentially change parameters of a black hole when
the
curvature at the horizon becomes comparable with Planckian curvature,
i.e. for
black holes of Planckian mass.
These remarks are of course  trivial. But if we exclude these evident
cases do
we still need quantum gravity to explain properties of macroscopic
black holes?

This question is not new. The standard derivation of Hawking quantum
radiation
of a black hole formally requires the integration over all (including
much
higher than Planckian)  frequencies of the initial
zero-point-fluctuations of a
quantum field. There is a belief that one can escape the formal usage
of
super-Planckian energies in the calculations. This point of view was
supported
by recent result by Unruh \cite{Unru:95}. He considered a model in
which due to
the presence of dispersion the frequencies of zero-point-fluctuations
are
restricted. Unruh  has shown  that nevertheless the  Hawking
radiation at late
time remains constant and  with high accuracy thermal.

A similar problem arises in connection of a black-hole entropy. We
saw, for
example,  that the statistical-mechanical entropy $\jj{S}^{SM}$ is
dependent on
the cut-off parameter. One might argue that such a cut-off must be
provided by
quantum gravity. For this reason $\jj{S}^{SM}$ is (at least
potentially) the
quantity which for its knowledge requires Planckian scale physics.
Does it mean
that the study of the thermodynamical properties might give us
information
about these scales? The above discussion indicates that it is
impossible. In
the standard gedanken experiments the observable quantity is
$\jj{S}^{TD}$.
$\jj{S}^{SM}$ (at least in the leading order) does not contribute to
$\jj{S}^{TD}$ and hence one cannot measure it. The
statistical-mechanical
entropy might be useful for description of excitations in the close
vicinity of
the horizon (for example of their damping). The main contribution to
$\jj{S}^{SM}$ is given by very high frequency modes  inside the
gravitational
barrier propagating very close to the horizon. It looks like that the
only
reasonable way to measure $\jj{S}^{SM}$ is to excite these modes. For
example,
one can do it by    colliding  particles of superhigh energy near the
horizon.
This experiment requires very high (super-Planckian) energies. But
having these
energies available  one can use them for study Planckian physics in
usual
Minkowski spacetime without any black holes.

To conclude, quantum gravity is required for understanding very
fundamental
problems of black holes, such as the problem of final state, but it
also looks
like that the thermodynamics of macroscopical black hole does not
provide us
with any new powerful tools for verifying the theory of  quantum
gravity.

\ensec{Acknowledgements}
This work was supported by the Natural Sciences and Engineering
Research
Council of Canada.

\end{document}